\documentstyle[twocolumn,prb,aps,epsf]{revtex}
\def\inseps#1#2{\def\epsfsize##1##2{#2##1} \centerline{\epsfbox{#1}}}
\begin{document}
\draft
\input{psfig}
\title{Spinodal decomposition to a lamellar phase: effects of
hydrodynamic flow}
\author{G. Gonnella[\onlinecite{BEPPE}], E. Orlandini[\onlinecite{ENZO}], 
and J.M. Yeomans}
\address{Theoretical Physics, Oxford University,
1 Keble Rd. Oxford OX1 3NP, UK}
\date{\today}
\maketitle
\begin{abstract}
Results are presented for the kinetics of domain growth of a
two-dimensional fluid quenched from a disordered to a lamellar
phase. At early times when a Lifshitz-Slyozov mechanism is operative
the growth process proceeds logarithmically in time to a frozen state
with locked-in defects. However when hydrodynamic modes become
important, or the fluid is subjected to shear, 
the frustration of the system is alleviated and the size
and orientation of the lamellae attain their equilibrium values.
\end{abstract}


The kinetics of the growth of ordered phases as a system is quenched
from a one- to a two-phase region provides a wealth of interesting
physical questions[\onlinecite{B94}]. Our aim here is to present results 
for phase separation in a two-dimensional
binary fluid in the case where equilibrium corresponds 
to a lamellar structure where the two coexisting phases arrange themselves in
stripes. We
consider {\em both} the early-time regime, where the growth is driven
by a Lifshitz-Slyozov evaporation-condensation
mechanism[\onlinecite{LS61}], 
{\em and} 
later times (or equivalently lower viscosities)
where hydrodynamic modes alter the kinetic behaviour. Our main
conclusions are that when the former mechanism is operative the
lamellae initially grow logarithmically but then
become frozen in a tangled configuration. It is only when hydrodynamic
movement becomes effective that topological defects can be removed
allowing the lamellae to line up and attain their correct equilibrium
width.
We show that the frustration of the system can also be alleviated by
subjecting it to shear.

Our results are relevant to a wide variety of physical systems where
competing interactions result in stable lamellar phases. One example
is a di-block copolymer melt comprising A- and B- type chains
covalently bonded end-to-end in pairs. If the A--B interaction is
repulsive the tendency towards bulk phase separation at low
temperatures is halted by the covalent bond and a
lamellar phase can be stabilised[\onlinecite{B91}]. 
In microemulsions similar structures result
from the presence of surfactant at the (say) oil--water interface which
leads to competition between a negative surface tension
and positive interface curvature free energy[\onlinecite{GS94}]. 
Other examples include dipolar fluids with long-range Coulombic
interactions[\onlinecite{SD93}] 
and chemically reactive binary mixtures where the ordering
tendency of the two components is counterbalanced by the mixing effect
of the reaction[\onlinecite{GC94}].

The problem is addressed using lattice Boltzmann 
simulations[\onlinecite{BS92}] of the
Navier-Stokes and convection-diffusion equations for a binary
fluid. 
These simulations follow the behaviour of a set of distribution
functions, related to the fluid densities and velocity, which evolve
according to a discretised Boltzmann equation. The densities and
momenta are conserved at each step of the simulation thus
approximating the continuum equations of fluid flow.

The technique has
two advantages which are particularly relevant here. Firstly it allows
simulations on long enough time scales to access hydrodynamic
behaviour. Secondly by using an implementation of the method described
by Orlandini {\em et.\ al.}[\onlinecite{OS95}] it is possible to 
introduce a free energy into
the model such that the fluid relaxes to an equilibrium state
determined by the minimisation of the input free energy. Hence we are
able to accurately control the equilibrium state that the fluid is
trying to reach.

The numerical method is described in detail in reference
[\onlinecite{OS95}]. Here we consider a model which
corresponds to an equilibrium free energy
density[\onlinecite{B75}]
\begin{equation}
{\cal F}\{\varphi\} = \int d^d x 
\{\frac{a}{2} \varphi^2 + \frac{b}{4} \varphi^4 
+ \frac{\kappa}{2} \mid \nabla \varphi \mid^2 +\frac {c}{2} (\nabla^2
\varphi)^2\}
\label{eqn1}
\end{equation}
where $\varphi$ is a scalar order parameter describing the concentration
difference between the two binary components. We take $b,c>0$ to
ensure stability. For $a>0$ the fluid is disordered; for $a<0$ it
prefers an ordered state. The nature of this depends on the value of
the surface tension.
For $\kappa$ sufficiently large for fixed $c$
the surface tension is positive and the fluid prefers to phase
separate into two homogeneous phases. As $\kappa$ is decreased,
however, at some $\kappa=\kappa_c\sim-0.8$ for $c=1$ 
the surface tension becomes
negative and it becomes favourable to introduce interfaces into the
system. The ordered state is now a lamellar or striped phase
with spacing determined by the competition between the free energy
gain due to interface creation and the repulsive force between
interfaces.
We shall compare the behaviour of the fluid under
sudden quenches from the disordered to the ordered phases for $\kappa$
greater than and less than $\kappa_c$.

Before presenting our results it is useful to summarise the known
behaviour of a binary fluid quenched into the ordered homogeneous
phase[\onlinecite{B94}]. 
Once domains of the two phases are well established
experimental and numerical data indicates that the growth is
self-similar with a typical length scale $R$ scaling with time as
$R\sim t^\alpha$. The exponent $\alpha$ is universal depending only on
a small number of parameters; the dimensionality, conservation laws
applicable to the order parameter and the presence or absence of hydrodynamic
degrees of freedom. Three growth mechanisms are believed to be
operative in two dimensions: (i) Lifshitz-Slyozov evaporation
of particles from interfaces of high curvature to recondense on
interfaces of lower curvature with $\alpha=1/3$,
(ii) a regime with $\alpha=1/2$ which has been explained as being due to
Brownian diffusion of interfaces and droplets[\onlinecite{SG85}], 
(iii) inertial hydrodynamic flow with $\alpha =
2/3$[\onlinecite{F85}]. 
Previous lattice
Boltzmann simulations[\onlinecite{OO95}] 
have shown a crossover from (i) to (iii) with
decreasing viscosity with (ii) not seen as might be expected because of the
lack of noise in the simulations. Lattice gas simulations, which do
include noise, show a crossover from (ii) to (iii)[\onlinecite{EC96}]. However 
puzzles do remain. For example, a simulation by Lookman {\em et. al.}[\onlinecite{LW96}] 
using a Ginzburg-Landau approach, shows a crossover from (ii) to (iii)
independent of whether or not noise is present.

\begin{figure}
\vskip -3mm
\inseps{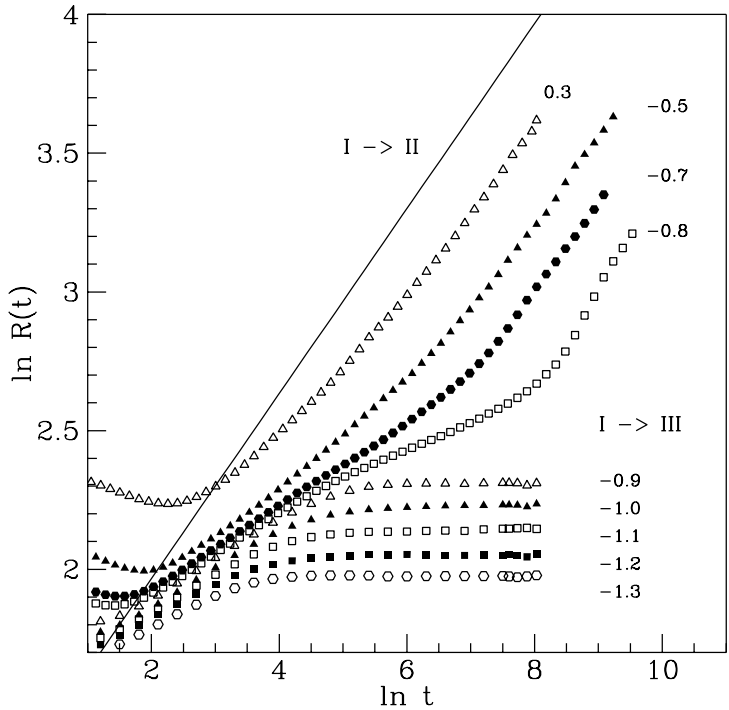}{0.5}
\vskip +5mm
\caption{Logarithm of a typical domain size as a function of the
logarithm of the time for different values of the surface tension and a
sufficiently high viscosity that the growth proceeds by the
Lifshitz-Slyozov mechanism. The straight line corresponds to a growth
exponent $\alpha=1/3$. For $\kappa$ sufficiently negative the domain
size grows logarithmically and then saturates as the system reaches a
frozen state of tangled lamellae.}
\end{figure}

Results for the evolution of the domain size $R(t)$ with time
following a quench to the ordered phase $a<0$ for different values of
$\kappa$ are shown in Fig. 1.
$R(t)$ was calculated as the first moment of the
circularly averaged structure factor. 
These simulations were run with high values of the viscosity where the
diffusive Lifschitz-Slyozov growth mechanism is expected to be
operative.

For $\kappa \ge  -0.8$, where the surface tension is positive, all
systems reach the scaling regime (i) after a sufficiently long
time. As $\kappa$ is decreased the domains take longer to reach a
given size. This is as expected because the driving force for
the Lifshitz-Slyozov growth, provided 
by the surface tension,
is lower. It is interesting to note that the scaling regime sets in at
approximately the same domain size in each of the runs. The data
presented in Fig. 1 is from a single run on a lattice of size $128
\times 128$ as our aim here is to
provide a qualitative comparison of the results as $\kappa$
is decreased
rather than the best quantitative value for $\alpha$ which has been
done elsewhere[\onlinecite{OO95}].

For $\kappa <  \kappa_c$ where the system
prefers to order in a lamellar phase 
a significantly different behaviour is
apparent. After initial transients there is a region of logarithmic
growth which corresponds to formation of the lamellae. The growth then
slows down and the fluid becomes frozen.
A similar pattern of growth has been observed for quenches into
microemulsion phases[\onlinecite{LM94,EC96}] and for systems 
with quenched defects[\onlinecite{GH95}]. 

\begin{figure}
\vskip -23mm
\inseps{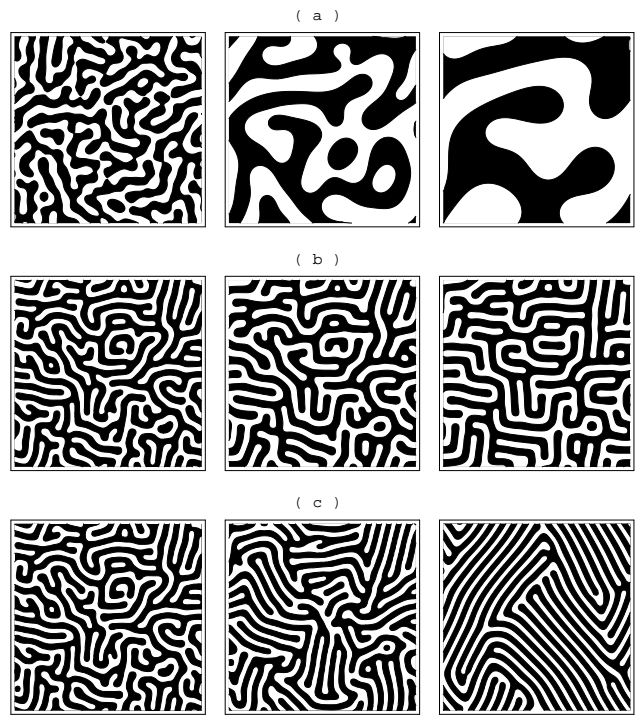}{0.6}
\vskip -18mm
\caption{Snapshots of the growth of domains with time. Each line,
reading from left to right, represents a different physical situation: (a)
a quench to the homogeneous two phase region $\kappa > \kappa_c$; 
(b) a quench to the
lamellar phase $\kappa < \kappa_c$ in a high viscosity fluid. The
lamellae form in a tangled pattern which becomes frozen in time; 
(c) a quench to the same final
parameters as (b) but for a low viscosity fluid. 
Hydrodynamic modes allow the lamellae to reorder giving, locally,
well-defined striped regions.} 
\end{figure}

A pictorial comparison of the domain growth for $\kappa > \kappa_c$
and $\kappa < \kappa_c$ is made in Figs. 2(a) and (b). In the former
case the system prefers to form elongated domains because of the
curvature term in the free energy. However this does not impede growth
into two homogeneous phases and the lamellae grow and become more
isotropic throughout the simulation. For $\kappa <\kappa_c$,
however, (Fig. 2b), the lamellae form in a tangled pattern which
becomes frozen in time.

We now turn to a  primary aim of this Letter 
which is to report the effect of hydrodynamic
modes on the quench to the lamellar state. To investigate this we ran
the simulations for $\kappa= -0.9$ for a low value of the
viscosity. A pictorial comparison of the domain growth
in the long time regime  for the high and low viscosity cases are given 
in Figs. 2b and 2c
respectively. 
It is apparent that in the latter case it becomes easier for the
lamellae to grow and shrink along their length thus allowing a
decrease in the frustration in the
system. This is because the long-range nature of the hydrodynamic
modes are allowing the fluid to reorder in such a way as to remove
topological defects from the fluid.
For the latest accessible times some defects remain although
the system is locally lamellar. 

\begin{figure}
\vskip -3mm
\inseps{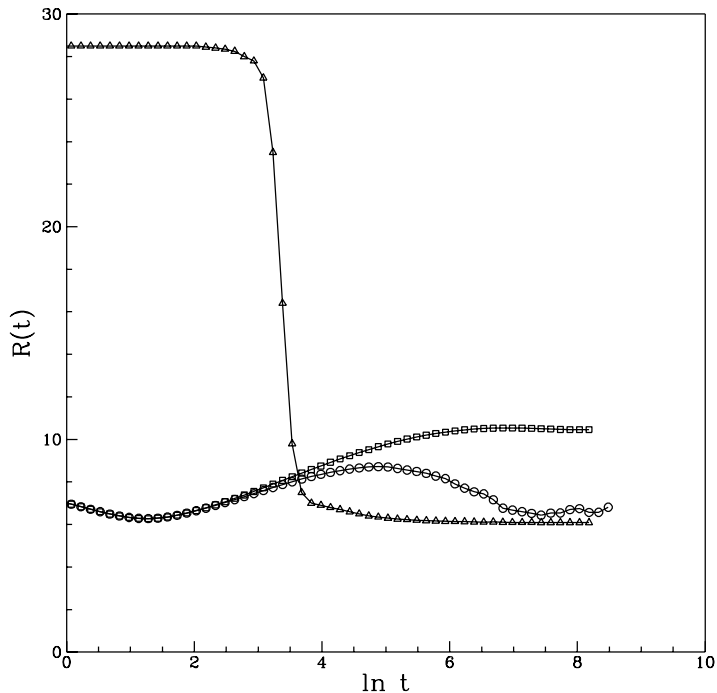}{0.5}
\vskip +5mm
\caption{Typical domain size as a function of the
logarithm of the time for a quench to the lamellar phase for $\Box$
random initial conditions and a high viscosity; $\bigcirc$ random initial
conditions and a low viscosity; $\triangle$ a sine wave as initial condition
and a high viscosity. In the second and third cases, for long times, the fluid
is not frustrated and the lamellae can attain their equilibrium width.}
\end{figure}

The rearrangements of the lamellae also have the effect of
allowing them to attain their equilibrium width. The growth of the
domain size with time following a quench to the lamellar phase for
high and low viscosities is compared in Fig. 3. In the latter case
there is a relatively rapid decrease in the domain width as
hydrodynamic modes become effective. As evidence that equilibrium is
indeed being achieved we also show in Fig. 3 a quench for the same
(high viscosity) parameters but with a one-dimensional sine wave as
the initial condition. This allows the lamellae to grow easily in an
ordered pattern. The final domain size achieved is very similar to
that obtained after hydrodynamic rearrangements of the frozen state.

As a further check on the mechanism causing the unlocking of the
frustrated lamellar state we simulated the behaviour of simple
topological defects. Examples are a broken lamella or an extra portion
of lamella inserted in an otherwise regular pattern. In each case
diffusive growth was only able to cause small changes of interface
shape whereas hydrodynamic rearrangements succeeded in removing the
defect from the fluid.

\begin{figure}
\vskip -30mm
\inseps{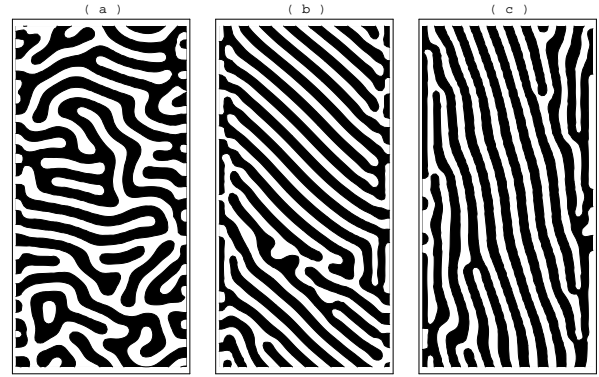}{0.5}
\vskip -30mm
\caption{Configurations of a lamellar fluid when a
shear flow is applied. The shear velocity is:
(a) v=0, (b) v=0.1 and (c) v=0.3.}
\end{figure}

Finally we consider another mechanism which allows topological disorder
in the lamellar state to be removed, the introduction of a
shear. 
To simulate the shear we imposed rigid walls moving with velocity $\pm
v$ on the left and right of the simulation area. A natural way to
implement rigid walls is to use bounce-back conditions at the boundary
sites[\onlinecite{CH91}]. The other edges of the system were
linked with periodic boundary conditions.

The fluid was quenched into the lamellar state 
giving the disordered configuration shown in Fig. 4(a). As the shear
was applied ordered lamellae quickly 
formed, aligned at an
angle to the direction of shear, as shown in Figure 4(b).
A further increase in the shear rate resulted in a decrease in the
angle between the direction of the lamellae and that of the shear 
(Figure 4(c)). We believe that this angle is determined by a
balance between the force on the lamellae due to the shear and the
elastic force which results because they are compressed. Presumably on
much longer time scales lamellae could be lost from the system,
releasing the elastic energy and allowing the lamellae to turn towards
the shear direction.

Thermal fluctuations are 
not included in the lattice Boltzmann simulations and it is
important to ask whether noise could be responsible for unpinning the
metastable state before the onset of hydrodynamics. To test this we ran
a simulation of Langevin dynamics for the free energy (\ref{eqn1}).
Tangled lamellar were formed in the same way as in the lattice
Boltzmann simulations suggesting that noise is ineffective in
bringing the system to a global thermodynamic equilibrium. This
conclusion is supported by cell dynamical
calculations[\onlinecite{BO90}] 
but disagrees
with results based on a time-dependent 
Ginzburg-Landau simulation[\onlinecite{PD95}].

To conclude, we have studied domain growth in a fluid with a
well-characterised, lamellar, equilibrium state. Diffusive growth
leads to a frozen pattern of tangled lamellae. The topological defects
in the fluid can be removed by hydrodynamic modes or applied shear. It
would be interesting to extend this work to three dimensions where a
new physical mechanism, the Rayleigh instability of tubes, is expected
to be operative[\onlinecite{S79}]. However such calculations would be extremely
demanding computationally.

~\\
\noindent ACKNOWLEDGEMENTS:\\
We thank F. Alexander, P. Coveney, A. Emerton, M. Swift and A. Wagner
for helpful discussions. GG and JMY acknowledge support from the EPSRC
and EO from an EU Human Capital and Mobility Fellowship.

\end{document}